# First principles studies of the size and shape effects on reactivity of the Se modified Ru nanoparticles

Sebastian Zuluaga and Sergey Stolbov

Physics Department University of Central Florida. Orlando, FL, USA

**Abstract.** We present here the results of our density-functional-theory-based calculations of the electronic and geometric structures and energetics of Se and O adsorption on Ru 93- and 105-atom nanoparticles. These studies have been inspired by the fact that Se/Ru nanoparticles are considered promising electrocatalysts for the oxygen reduction reaction (ORR) on the direct methanol fuel cell cathodes and the oxygen binding energy is a descriptor for the catalyst activity towards this reaction. We find the character of chemical bonding of Se on a flat nanoparticle facet to be ionic, similar to that obtained earlier for the Se/Ru(0001) surface, while in the case of a low coordinated Ru configuration there is an indication of some covalent contribution to the bonding leading to an increase in Se binding energy. Se and O co-adsorbed on the flat facet, both accept electronic charge from Ru, whereas the adsorption on low-coordinated sites causes more complicated valence charge redistribution. The Se modification of the Ru particles leads to weakening of the oxygen bonding to the particle. However, overall, O binding energies are found to be higher for the particles than for Se/Ru(0001). High reactivity of the Se/Ru nanoparticles found in this work is not favorable for ORR. We thus expect that larger particles with well-developed flat facets are more efficient ORR catalysts than small nanoparticles with a large fraction of under-coordinated adsorption sites.

# 1. Introduction

The proton exchange membrane fuel cells (PEMFC) and the direct methanol fuel cells (DMFC), as clean renewable sources of energy, can offer great advantages for various applications. However, the Pt-based catalysts, used in both electrodes of the fuel cell, make these devices unacceptably expensive. Furthermore, performance of both PEMFC and DMFC suffers from low rate of the oxygen reduction reaction (ORR) on the Pt (or Pt-based) cathode. This is thus not surprising that a significant effort has been made in search for new cost-effective Pt-free materials that efficiently catalyze ORR.

One of the focuses of the search is on the Se-Ru system. It has been more than two decades since Alonso-Vante and Tributsch reported a relatively high ORR rate for $Mo_{4.2}Ru_{1.8}Se_8$ [1]. More recent studies show that the best electrocatalytic performance is achieved with the $Ru_xSe_y$ nanoparticles [2-11]. The main conclusions drawn from these works are that: a) the best catalytic activity is achieved for the few nanometer particles with the Ru core covered with Se at composition $Ru_{85}Se_{15}$, b) The Ru core has the bulk-like hcp structure, c) it is not clear how Se is distributed over the particle surface. Recent calculations [12] show that Se adsorbates with sub-monolayer coverage accept electronic charge from Ru and, therefore, repel each other. The experimental data on the electro-chemical nuclear magnetic resonance and X-ray photoelectron spectra also suggest the electron charge transfer from Ru to Se [13]. As a result, Se atoms tend to scatter over the surface rather than from 2D islands or 3D structures [12,14].

The Se/Ru structures have very promising electrocatalytic properties including: a) relatively high ORR rate, b) excellent tolerance to methanol oxidation, c) high selectivity of four-electron oxygen reduction to water (small fraction of $H_2O_2$), and d) high stability towards oxidation. These catalysts are extensively studied [15], however, further progress in this direction requires deeper understanding the factors controlling the ORR energetics on Se/Ru. For example, the mentioned above calculations [12] have been performed for the Ru(0001) surface. This is the lowest energy surface for the hcp structure. One can thus assume that in the large enough particles the (0001) facets will dominate. However, the nano-size effects on the Se distribution over the Ru core have still to be understood.

Another issue to address is the mechanisms underlying the improvement of the Ru surface activity towards ORR upon Se modification. It has been shown that the ORR activity of electrocatalyst surfaces can be evaluated based on knowledge of the free energy of the reaction intermediate states, in particular those of OOH, O and OH [16], whereas their free energies are mostly determined by the binding energy of the intermediate to the catalyst surface [16, 17]. Furthermore, for many materials, the binding energies of OOH ($E_B(OOH)$) and OH ($E_B(OH)$) are found to be approximately proportional to binding energy of atomic oxygen ($E_B(O)$) [17 - 20]. Therefore, $E_B(O)$ can be used as an ORR descriptor for preliminary screening of materials. It has also been shown that Ru(0001) surface is too reactive ($E_B(O)$ is too high) to efficiently catalyze ORR [21]. One thus can expect that the Se modification reduces $E_B(O)$ on the Ru surface. The strength of covalent O – metal bond is usually associated with hybridization of oxygen and metal states, which has been shown to depend on the metal *d*-band center position [22]. The $E_B(O)$ - *d*-band center correlation has been reported for various materials including the ORR catalysts [18, 23,24]. It is important to note, however, that the Se modification does not change noticeably the density of d-electronic states of Ru surface atoms [12], suggesting that alternative mechanisms of the Se effect on the Ru surface reactivity have to be explored. It has been shown recently [25] that both Se and O accept electronic charge from Ru upon adsorption and, hence, repel each other. This repulsion increases total energy of the system and thus leads to decrease in the O binding energy. However, again, these results were obtained for a selenium modified Ru(0001) surface, and it still needs to be understood how geometry of small Se/Ru nanoparticles affects this mechanism.

In the present work we address the questions raised above. We obtain from first principles calculations the formation energies of the Se sub-monolayer on Ru nano-particles as a function of the Se coverage, as well as oxygen binding energies for different configurations of Se on the Ru particles. The calculations have been performed for two Se/Ru particles of ~1.2 nm size. Such a choice allows us to study in details the effects of under-coordinated Ru sites on the energetics of Se and O adsorption and the electronic structure of the system. On the other hand, comparison of the results to those obtained for the flat Ru(0001) surface provides a clear suggestion of what energetics and electronic structure one can expect for the 2 nm – 5 nm particles, which have been observed in experiment and whose systematic first principles studies are still not feasible.

## 2. Computational details

In this work we study the Se/Ru particles with 105 Ru atom core. Some calculations were also performed for a 93 Ru atom particle with relatively large facets.

All calculations in this work have been performed using the VASP5.2 code [26] with projector augmented wave potentials [27] and the Perdew - Burke - Ernzerhof (PBE) version of the generalized gradient approximation (GGA) for the exchange and correlation functional [28].

To maintain periodicity, the particles were situated in a cubic vacuum box of 24 Å side. Since no dispersion of the electronic states occurs in such systems, the calculations were performed only for Γ point of the Brillouin zone, as it usually done for non-periodic systems modeled by a periodic computational method. The cut of energies of 400 eV and 600 eV were used for the plane wave expansion of wave functions and charge density, respectively. To achieve structural relaxation, the atomic positions were optimized to obtain equilibrium geometric structures in which forces acting on atoms do not exceed 0.02 eV/Å.

The binding energy of the adsorbate $X$ ($X$ = O, Se) on a Ru particle was calculated as:

$$E_b(X) = E_{tot}(cluster) + E_{tot}(X\ atom) - E_{tot}(Se\ /cluster) \qquad (1)$$

Similarly the formation energy per adsorbate, when more than one $X$ atom is adsorbed on the cluster, was defined as:

$$E_f(X) = [E_{tot}(cluster) + nE_{tot}(X\ atom) - E_{tot}(nSe\ /cluster)]/n \qquad (2)$$

where $n$ is the number of $X$ atoms adsorbed on the surface. These energies are thus positive if adsorption is favorable. Since there are various non-equivalent adsorbtion sites on the Ru core, the formation energy per adsorbate characterizes the average binding energy of the adsorbate on the core.

The valence electron charge redistribution upon O and/or Se adsorption was defined as follows:

$\delta\rho(\mathbf{r}) = \rho_{SeRu}(\mathbf{r}) - \rho_{Ru}(\mathbf{r}) - \rho_{Se\text{-}atom}(\mathbf{r}) -$ (redistribution upon Se adsorption),

$\delta\rho(\mathbf{r}) = \rho_{OSeRu}(\mathbf{r}) - \rho_{Ru}(\mathbf{r}) - \rho_{O\text{-}atom}(\mathbf{r}) - \rho_{Se\text{-}atom}(\mathbf{r}) -$ (redistribution upon Se and O adsorption), where the right-side equation terms denote the valence charge densities of the Ru surface adsorbed with O and/or Se, clean Ru surface, isolated O atom, and isolated Se atom, respectively. The two-dimensional cuts of $\delta\rho(\mathbf{r})$ shown in this article are plotted for the density interval ±0.022 e/Å. The uniform blue and red regions seen in these cuts thus correspond to $\delta\rho(\mathbf{r})$ values lower than -0.022 e/Å or higher than 0.022 e/Å, respectively.

Geometric structures of the system under consideration and two-dimensional cuts of $\delta\rho(\mathbf{r})$ shown in this article have been plotted using the Xcrysden software [29].

## 3. Results and Discussion

First we calculated the optimized geometry of the 105 atom Ru core. The initial shape of the cluster was chosen as a piece of the hcp bulk Ru. To avoid an unphysical force cancelation due to symmetry of the system, we have randomly deviated atoms from the equilibrium by 0.02 Å. The relaxed Ru core cluster is found to retain an hcp-like structure, although the optimization changes some bond length noticeably. It is important for our further studies that the chosen core cluster has various under-coordinated Ru sites which represent typical geometry of small nanoparticles (see Fig. 1). As seen from Fig. 1, the binding energy of Ru atoms to the cluster depends on local geometry and, in particular, on coordination numbers. Nevertheless, all surface Ru atoms are found to bind to the cluster strong enough to ensure its stability.

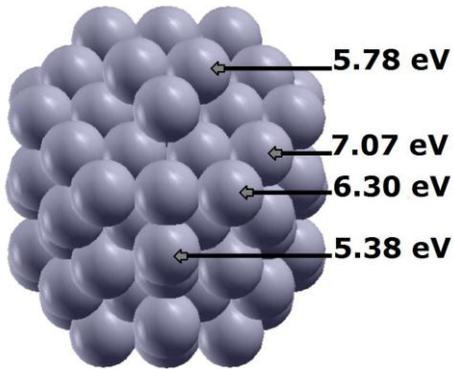

Fig. 1. Optimized geometric structure of the 105-atom Ru cluster. The inserted numbers provide binding energies for some Ru sites.

### 3.1. Energetics of formation of Se/Ru core-shell structures

In order to understand the mechanism underlying formation of the Se shell on the Ru core, we obtained the optimized geometries, electronic structures and Se formation energies for three Se/Ru structures (see Fig. 2) with sub-monolayer Se coverage (1, 18, 30 and 40 Se atoms on the Ru 105 atom cluster). These structures approximately correspond to the 1.6 30, 50, and 65% Se coverage, respectively. The calculated average formation energies per Se atom $E_f(Se)/n$ are plotted versus Se coverage in Fig. 3. The $E_f(Se)/n$ values obtained for the flat Ru(0001) surface are also provided for the reference. One can see that $E_f(Se)/n$ are decreasing with increase in the Se coverage. The coverage dependence of $E_f(Se)/n$ is found to be similar to that obtained for Se on Ru(0001). However, the absolute values of $E_f(Se)/n$ are higher on the Ru cluster than on Ru(0001). It has been shown for Ru(0001) [12] that Se atoms accept the electronic charge from Ru upon adsorption and repel each other that leads to a decrease in $E_f(Se)/n$ upon an increase in Se coverage. To understand the similarities and differences in Se adsorption on the flat Ru surface and Ru cluster, we calculated the valence charge density redistribution upon adsorption of two Se atoms on a flat facet of the Ru 93 atom cluster and around an apex of the Ru 105 atom cluster (see Fig. 4). It is clearly seen for the flat facet configuration that Se accepts the valence charge from Ru and the electronic charges accumulated on the neighboring Se atoms repel. The repulsion causes an increase in the total energy of the system and, hence, results in reduction of the Se binding energy. This effect is thus similar to that revealed for Se on Ru(0001) [12]. However, if Se atoms are adsorbed around an under-coordinated Ru apex, character of the Se – Ru bonding is found to be more complicated. As shown in Fig. 4b, the electronic charge transfer to Se is not so pronounced. Furthermore, there is some valence charge accumulation in the middle of the Se – Ru bonds suggesting a covalent contribution to the bonding, in addition to the ionic one. More details on the character of Se – Ru bonding can be derived from analysis of the local densities of electronic states (LDOS) plotted in Fig. 5. As seen from the figure, adsorption of Se on the flat Ru facet doesn't change the densities of Ru d-states noticeably, while Se adsorbed around the Ru apex causes widening of the Ru d-band due to hybridization between the Se p- and Ru d-states. Furthermore, the Ru apex atoms located between Se atoms screen the repulsion. As a result, the Se binding energy (5.45 eV) in this configuration is higher than that on Ru(0001). For example, for a similar configuration – three Se atoms situated at the neighboring hcp Ru(0001) sites – the Se binding energy is 5.21 eV [12]. We thus may expect non-uniform distribution of Se atoms over small Ru cluster surface with some accumulation of Se around under-coordinated Ru sites, while on the flat Ru(0001) surface the Se - Se repulsion is found to lead to a homogeneous distribution of Se atoms [12].

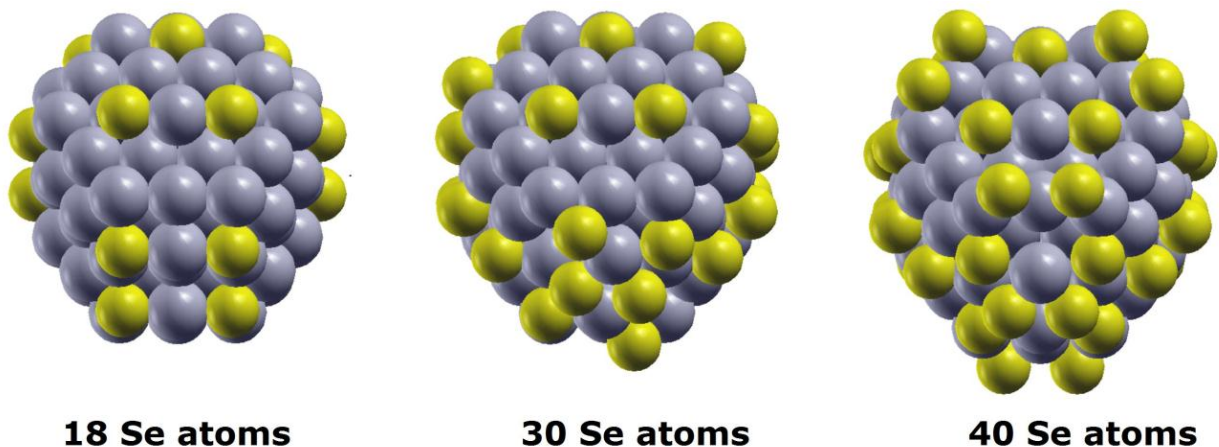

Fig. 2. Geometric structures calculated for three different Se coverages on the Ru 105-atom cluster.

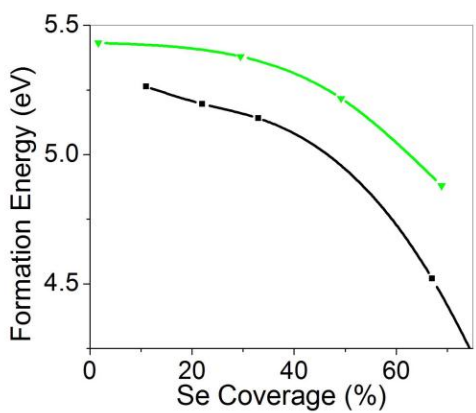

Fig. 3. Formation energy per Se atom as a function of the Se coverage on the Ru 105-atom cluster (green line) and on the Ru(0001) surface (black line).

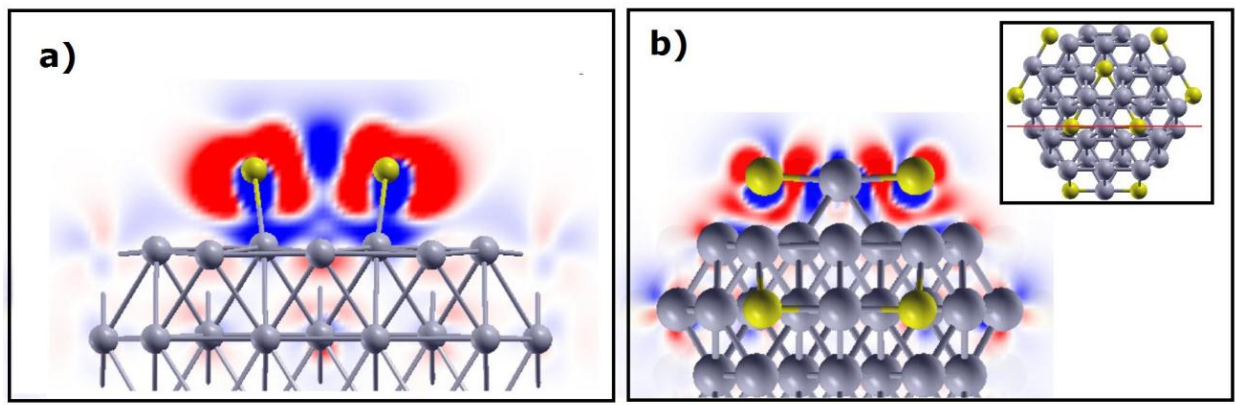

Fig. 4. Electronic charge density redistribution upon Se adsorption on a flat facet of the Ru 93-atom cluster (panel a) and around the apex of the Ru 105-atom cluster (panel b). Insert in panel b shows the projection of the cut plane for $\delta\rho(\mathbf{r})$.

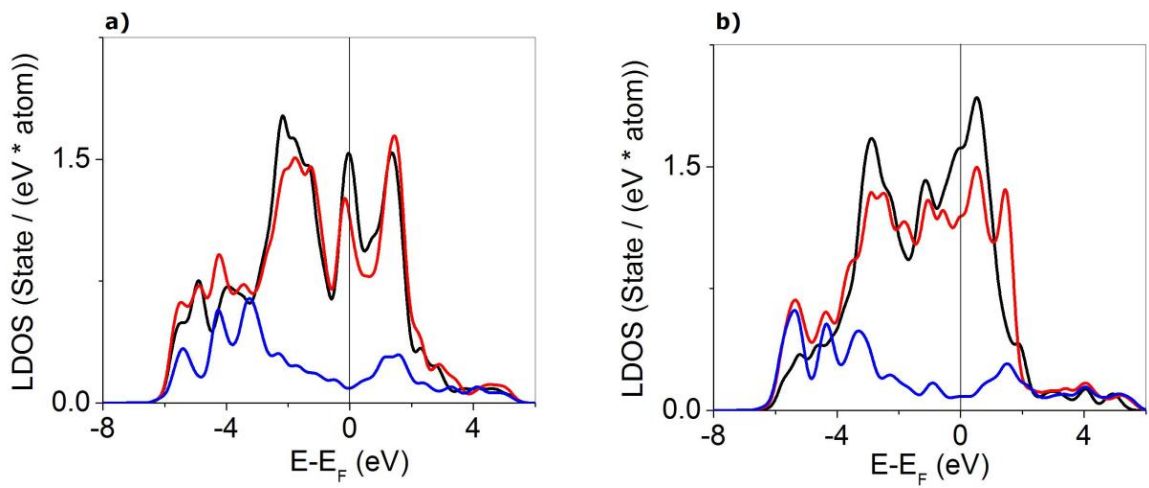

Fig. 5. LDOS calculated for the flat facet in the Ru 93-atom cluster (left panel) and for the apex of the Ru 105-atom cluster (right panel). The black and red lines represent the densities of the Ru d-states calculated for the clean and Se-adsorbed clusters, respectively, while the blue lines represents the Se p-states.

*3.2. Oxygen adsorption on the Se/Ru nanoparticle*

Since our goal is to evaluate the nanoparticle size and shape effects on ORR on the Se/Ru system, and the oxygen binding energy is considered to be an ORR activity descriptor [16,17], we calculate $E_b(O)$ for a number of non-equivalent adsorption sites on the clean Ru 105 atom core and that pre-adsorbed with 18 Se atoms (see Fig. 6) that corresponds to about 30% Se coverage. The calculation results are listed in the Table 1. One can see that for the clean Ru core the $E_b(O)$ are in a wide range of 5.91 – 6.53 eV, which is determined by a variety of local geometries including coordination number of Ru atoms and number of O – Ru bonds. For example, the highest $E_b(O)$ are obtained for the adsorption sites #1 and #5, for which O makes three bonds with low-coordinated Ru atoms, while oxygen adsorption on the sites # 2 and #3 with higher-coordinated Ru atoms results in the lowest $E_b(O)$ values. It is important to note that, in general, the Ru 105 cluster is found to be more reactive than the Ru(0001) surface, for which $E_b(O) = 6.25$ and 5.82 eV for the hcp and fcc adsorption sites, respectively [25].

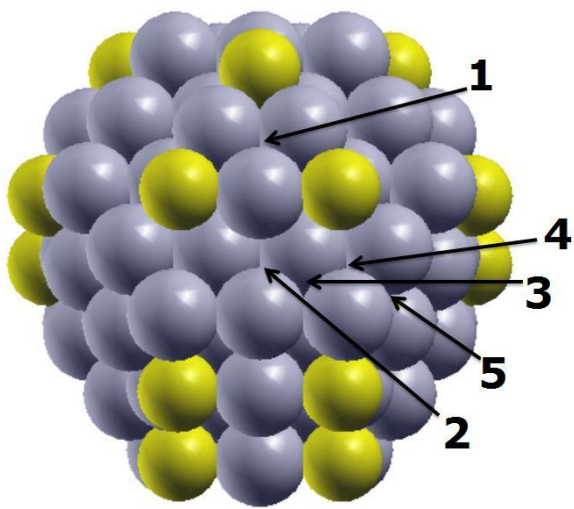

Fig. 6. Oxygen adsorption sites on the Ru 105-atom cluster considered in this work.

As seen from the Table 1, presence of Se causes a reduction in the oxygen binding energies on the Ru cluster. The $E_b(O)$ values are found to be in a wide range of 5.27 – 6.00 eV on the Se modified Ru cluster. The obtained binding energies correlate with the O – Se distances and the number of Se atoms around the adsorbed O. For example, presence of three neighboring Se decreases $E_b(O)$ at the #1 site by 1.26 eV, while one Se neighbor of approximately the same O – Se distance (site #3) causes the $E_b(O)$ reduction only by 0.36 eV. On the other hand, one Se neighbor, located closer to O, reduces $E_b(O)$ at the #4 site by 0.87 eV. If we compare the obtained $E_b(O)$ to those on Se/Ru(0001), we can conclude that, in general, the small Se modified Ru particle is more reactive than the flat Se/Ru(0001). Indeed, for Ru(0001) covered with 1/3 monolayer of Se (close the Se coverage on the Ru cluster considered above), $E_b(O) = 5.31$ eV for O adsorbed on an hcp site [25], which is smaller than $E_b(O)$ obtained for most sites on the particle. As concluded in Ref. 25, reactivity of the Se/Ru(0001) surface is still higher than optimal one for ORR. We find here that the considered Se/Ru particle with a significant number of under-coordinated Ru sites is even more reactive than Se/Ru(0001). This finding suggests that large Se/Ru nanoparticles with well-developed flat facets are more favorable for ORR than small Se/Ru clusters.

In order to trace the obtained Se effect on $E_b(O)$ to the electronic structure of the system, first, we come back to Fig. 5 to compare LDOS of Ru in the Se co-adsorbed flat facet and apex. We find that, in the case of the flat facet, Se adsorption does not change noticeably the densities of the d-electronic states of Ru. We can thus conclude that, like for the flat Se/Ru(0001), the Se effect on $E_b(O)$ is not caused by the change in the Ru LDOS due to hybridization between Se p- and Ru d-states. However, Se adsorbed around the apex causes a decrease in Ru LDOS around the Fermi-level that may lead to the above-mentioned significant reduction in the binding energy of oxygen adsorbed on the apex top.

For further understanding of the nature of the Se effect on oxygen bonding to Ru, we analyze charge density redistribution upon Se and O adsorption on the Ru cluster. As shown in Fig. 7a, if Se and O are adsorbed on a flat facet, both species accept a significant amount of the electronic charge from Ru, which causes Se – O repulsion. A wide blue area around Ru surface atoms seen in the figure suggests that the charge density is taken from delocalized s- and p-states of Ru, while the blue region between Se and O reflect localization of the electronic density on Se and O atoms due to electrostatic repulsion. We thus find that in the case of the flat Ru facet the mechanism of reduction of $E_b(O)$ on Ru upon Se co-adsorption is the same as for Se/Ru(0001) [25] – electrostatic repulsion. Meanwhile, oxygen adsorption on the Se modified Ru apex causes charge transfer from Se to O (see Fig 7b). The apex with the lowest Ru coordination in the particle considered here provides the "extreme" geometry for Se and O bonding. One thus may expect that, for more compact configurations, the Se – Ru and O – Ru bonding are more similar to that obtained for the flat facet.

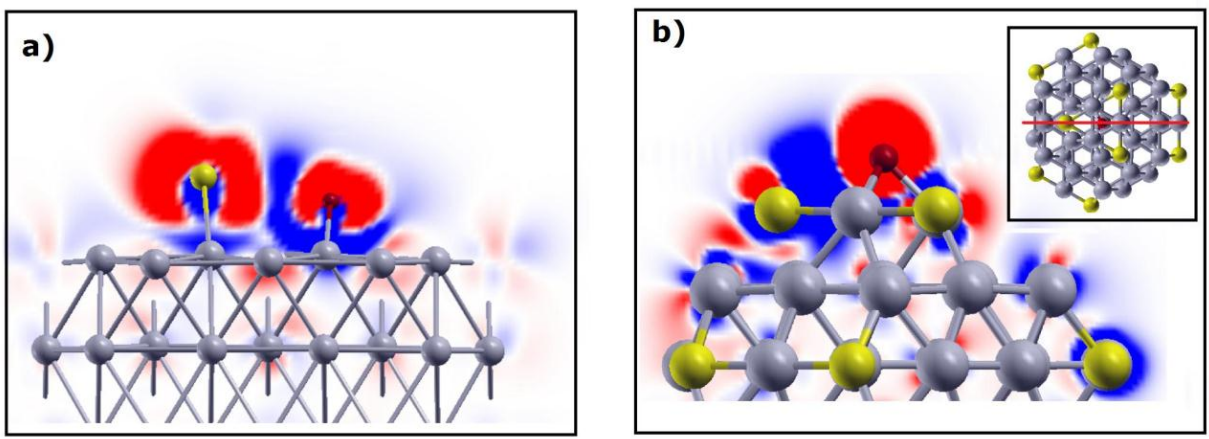

Fig. 7. Electronic charge density redistribution upon Se and O adsorption on a flat facet of the Ru 93-atom cluster (panel a) and around the apex of the Ru 105-atom cluster (panel b).

4. **Conclusions**

We have performed DFT-based calculations of the electronic and geometric structures and energetics of the Se and O adsorption on the Ru 105- and 93-atom clusters. The low-coordinated Ru-sites in the clusters are found to be more reactive than those in flat facets.

Character of Se – Ru and O - Ru chemical bonding and binding energies of Se and O are found to depend strongly on the Ru coordination and presence of co-adsorbed species. The Se atoms make purely ionic bonds to the flat Ru facet, similar to that reported for the Se/Ru(0001) system [12,25], while a mixed ionic-covalent Se – Ru bonding are found to take place for the low coordinated Ru apex configuration. Enhanced reactivity of small Se-modified Ru particles is not favorable for the oxygen reduction reaction. Therefore larger particles with well-developed facets are suggested to be more efficient ORR catalysts than small particles with a large fraction of low-coordinated sites.

Table 1. Se effects on oxygen adsorption on the Ru nanoparticle

| Adsorption site | # of O – Ru bonds | O – Ru bond lengths (A) | # of Se neighbors | Distance to Se atoms (A) | $E_b(O)$ on clean Ru cluster (eV) | $E_b(O)$ on Se/Ru cluster (eV) |
|---|---|---|---|---|---|---|
| 1 | 3 | 2.10 | 3 | 3.26 | 6.53 | 5.27 |
| 2 | 2 | 1.91 / 2.08 | 2 | 3.41 | 5.91 | 5.39 |
| 3 | 3 | 2.00 / 2.00 / 2.10 | 1 | 3.23 | 5.95 | 5.59 |
| 4 | 2 | 1.97 / 2.00 | 1 | 2.91 | 6.47 | 5.60 |
| 5 | 3 | 2.07/ 2.05 / 2.04 | 1 | 3.58 | 6.53 | 6.00 |

**Figure Captions**

Fig. 1. Optimized geometric structure of the 105-atom Ru cluster. The inserted numbers provide binding energies for some Ru sites.

Fig. 2. Geometric structures calculated for three different Se coverages on the Ru 105-atom cluster.

Fig. 3. Formation energy per Se atom as a function of the Se coverage on the Ru 105-atom cluster (green line) and on the Ru(0001) surface (black line).

Fig. 4. Electronic charge density redistribution upon Se adsorption on a flat facet of the Ru 93-atom cluster (panel a) and around the apex of the Ru 105-atom cluster (panel b). Insert in panel b shows the projection of the cut plane for $\delta\rho(\mathbf{r})$.

Fig. 5. LDOS calculated for the flat facet in the Ru 93-atom cluster (left panel) and for the apex of the Ru 105-atom cluster (right panel). The black and red lines represent the densities of the Ru

d-states calculated for the clean and Se-adsorbed clusters, respectively, while the blue lines represents the Se p-states.

Fig. 6. Oxygen adsorption sites on the Ru 105-atom cluster considered in this work.

Fig. 7. Electronic charge density redistribution upon Se and O adsorption on a flat facet of the Ru 93-atom cluster (panel a) and around the apex of the Ru 105-atom cluster (panel b).